\documentclass[aps,prl,twocolumn,groupedaddress,a4paper,twoside]{revtex4-1}

\usepackage{amsmath}
\usepackage{txfonts}

\usepackage[T1]{fontenc}
\usepackage{textcomp}

\usepackage[pdftex]{graphicx}

\usepackage{enumerate}

\begin{document}


\title{Simulation of Equilibrated States via Molecular Monte Carlo Method of Systems Connected to 3 Reservoirs}

\author{Yuki Norizoe}
\author{Toshihiro Kawakatsu}
\affiliation{Department of Physics, Tohoku University, 980-8578 Sendai, Japan}




\date{June 8, 2011}

\begin{abstract}
Metastable structures in macromolecular and colloidal systems are non-equilibrium states that often have long lifetimes and cause difficulties in simulating equilibrium. In order to escape from the long-lived metastable states, we propose a newly devised method, molecular Monte-Carlo simulation of systems connected to 3 reservoirs: chemical potential $\mu$, pressure $P$, and temperature $T$. One of these reservoirs is adjusted for the thermodynamic equilibrium condition according to Gibbs-Duhem equation, so that this adjusted 3rd reservoir does not thermodynamically affect phases and states. Additional degrees of freedom, \textit{i.e.} system volume $V$ and the number of particles $N$, reduce kinetic barriers of non-equilibrium states and facilitate quick equilibration. We show globally-anisotropic defect-free ordered structures, \textit{e.g.} string-like colloidal assembly, are obtained via our method.
\end{abstract}


\maketitle

Metastable states, \textit{e.g.} amorphous solids of colloids~\cite{Akcora:2009} and stalks in membrane fusion~\cite{Chernomordik:2008}, correspond to the regions where the free energy takes local minima in phase space. In conventional simulation of the canonical ensemble ($NVT$-ensemble), once the system is trapped in these metastable regions, the system tends to stay long over acceptable computational time in these non-equilibrium states. Even if the system is allowed to leave non-equilibrium to ordered equilibrium states, constant $N$ and $V$ cause defects in ordered structures. This situation breaks global anisotropy of the ordered structure and compels the periodicity of the structure to fit the system size. For finding defect-free ordered structures, $N$ and the rectangular system box size $\left( L_x, L_y, L_z \right)$ need fine tuning to both the anisotropy and the periodicity, which are not known \textit{a priori}.

In advanced simulation techniques, \textit{e.g.} multicanonical ensemble method~\cite{Berg:1991}, the whole phase space is nearly homogeneously sampled via artifitial weights that reduce occurrence probability of metastable states at constant $N$ and $\left( L_x, L_y, L_z \right)$. However, the equilibrium microstates obtained by these advanced techniques are limited to those with the given set of $N$ and $\left( L_x, L_y, L_z \right)$. Free energy landscapes at the same particle density, $N/V$, and different sets of $N$ and $\left( L_x, L_y, L_z \right)$ are not searched. These extensive variables need manual and simultaneous fine tuning for the sake of determining the most stable state over these landscapes, which is a challenging task. For example, for perfect crystals, both $N$ and $\left( L_x, L_y, L_z \right)$ must be integer multiples of the unit structure. In addition, prior to production runs, these techniques also require advanced programming and precise adjustments of the artifitial weights. Furthermore, unphysical sampling processes disallow tracing physical trajectories in the phase space.

In the present article, we devise and perform molecular Monte Carlo simulation of systems connected to the 3 reservoirs (3-reservoirs method), in order to study equilibrated colloidal assembly. We venture to connect the 3 reservoirs, though Gibbs-Duhem equation (GD eq.) limits the number of reservoirs to 2. Additional degrees of freedom, $N$ and $\left( L_x, L_y, L_z \right)$, are equivalent to the additional dimensions of the phase space and provide shortcuts, through the extended regions of the phase space, from the non-equilibrium to the equilibrium state. The system itself is allowed to make the spontaneous and simultaneous fine tuning of $N$ and $\left( L_x, L_y, L_z \right)$ for the ordered structure. Owing to physical sampling processes, the physical trajectories in the phase space are traced in simulation via our method.

Guggenheim formally introduced Boltzmann factor of the ensemble with the 3 reservoirs~\cite{Guggenheim:1939}. Later, Prigogine and Hill also analytically studied this ensemble~\cite{Prigogine:1950,Hill:StatisticalMechanicsPrinciplesAndSelectedApplications}. However, these early authors concentrated on formalism, \textit{i.e.} mathematical aspects of the partition function, since their goal was to obtain a universal and generalized formula for a partition function that is applicable to any thermodynamically acceptable ensembles~\cite{Sack:1959,Koper:1996}. Physical aspects of this ensemble were totally left for the future. Here we study the physical aspects of this ensemble, mostly in intuitive and thought experimental ways.

Despite GD eq., a system with the 3 reservoirs is readily constructed in experiments, for example, when a wall of a system box placed in the grand canonical ensemble is replaced with a free piston connected to another reservoir, \textit{i.e.} $P$. We theoretically build thermodynamics and statistics of the systems with the 3 reservoirs in the present article. We tentatively call this ensemble $\mu PT$-ensemble.

As an example of thermodynamics, we pack particles into a diathermal box with a free piston, placed in environment at constant $(T, P)$. These two intensive variables determine the other intensive variables of this system, \textit{e.g.} $\mu$, $\rho = N/V$, and free energy per particle. This means thermodynamic degrees of freedom of this system equal 2, which results from GD eq. In conventional simulation, $N$ is also fixed at some value ($NPT$-ensemble), whereas states and phases of the system are independent of $N$; $N$ only scales extensive variables. Briefly, phase diagrams constructed in $PT$-plane are independent of extensive variables. Rather than fixing insignificant $N$, we connect a reservoir of $\mu$, determined from $(T, P)$. This reservoir does not affect phases and states, provided $\mu$ and $(T, P)$ are in equilibrium. This condition corresponds to a thermodynamically stable point (TSP) based on GD eq., resulting in an equation of state linking $T, P$, and $\mu$. At TSP, extensive variables are freely scaled, \textit{i.e.} indeterminate and fluctuating, while all the intensive variables are kept. This means, in simulating this system, we can choose simulation runs at small $N$, which are computationally advantageous. Gibbs free energy per particle, which should be minimized in $NPT$-ensemble prior to adding the 3rd reservoir of constant $\mu$, is still unchanged even after this 3rd reservoir is added.

In the above example, $(T, P)$ is given from the outside of the system; $\mu$ is adjusted according to these $(T, P)$ as an additional reservoir. Two other combinations $(T, \mu)$ and $P$, and $(\mu, P)$ and $T$ also work. Gibbs free energy per particle, grand potential per volume, and the thermodynamic potential of $\mu PS$-ensemble per $S$, \textit{i.e.} $(E - \mu N + PV) / S$, are simultaneously minimized in $\mu PT$-ensemble, where $S$ and $E$ denote entropy and internal energy respectively. This simultaneous minimization of the 3 free energy densities is originated from GD eq. When we connect the 3rd reservoir to the system and set the system at TSP, prior to the connection, the corresponding 3rd intensive variable needs adjusting. Through this adjustment, the corresponding free energy density is minimized. The same system at the same TSP is also built by the two other combinations of the 3 intensive variables, which results in the simultaneous minimization of the two other free energy densities. On the other hand, in the other ensembles, any sets of corresponding 3 external parameters, \textit{e.g.} $(T, V, N)$ in $NVT$-ensemble, can be chosen without the adjustment. Therefore, only thermodynamic potential of the ensemble, \textit{e.g.} Helmholtz free energy $F$ in $NVT$-ensemble, is minimized.

Here we sketch a system with the 3 reservoirs. Reservoirs 1 and 2, composed of ideal particles with densities $\rho_{\text{res}}^{(1)}$ and $\rho_{\text{res}}^{(2)}$, specify $\mu$ and $P$ respectively. The system and these 2 reservoirs are connected to reservoir 3, a thermostat. As another example of thermodynamics, we consider a system consisting of ideal particles. At TSP, a relation, $\rho_{\text{res}}^{(1)} = \rho_{\text{res}}^{(2)} = \rho$, holds. However, when $\rho_{\text{res}}^{(1)} > \rho_{\text{res}}^{(2)}$, both $N$ and $V$ diverge, since reservoir 1 continues supplying particles and reservoir 2 supplying the volume. When $\rho_{\text{res}}^{(1)} < \rho_{\text{res}}^{(2)}$, both $N$ and $V$ vanish. Therefore, the system reaches equilibrium only at TSP. This also applies to systems of interacting particles. We utilize these divergence and vanishment as criteria for heuristically, \textit{e.g.} by bisection method, determining TSP in our simulation.

The detail of the statistical mechanics of 3-reservoirs method will be discussed in our forthcoming article~\cite{NorizoeForthcomingPaper:2010}. The partition function of $\mu PT$-ensemble is,
\begin{align}
\label{eq:SingleComponentPartitionFunctionmuPT}
	&\Omega ( T, P, \mu )  \notag \\
	&:= \frac{ P }{ k_B T } \int_{0}^{\infty} dV \sum_{N=0}^{\infty} \left\{ \exp \left[ \frac{ \mu }{ k_B T } N \right] \exp \left[ - \frac{ PV }{ k_B T } \right] Z_N(T, V) \right\},
\end{align}
where $k_B T$ denotes thermal energy and $Z_N(T, V)$ the partition function of the system in $NVT$-ensemble. The first prefactor of this equation, $P / k_B T$, comes from conventions of $NPT$-ensemble~\cite{Frenkel:UnderstandingMolecularSimulation2002}. $\Omega ( T, P, \mu )$ is obtained in a similar manner as for the partition functions of the grand canonical and $NPT$-ensembles~\cite{Frenkel:UnderstandingMolecularSimulation2002}, as natural extension of these two ensembles. Boltzmann factor defined from $\Omega ( T, P, \mu )$ is, at fixed $N$, consistent with Boltzmann factor of $NPT$-ensemble and, at fixed $V$, Boltzmann factor of the grand canonical ensemble.

3-reservoirs simulation method is constructed based on conventional Monte Carlo simulation methods in the grand canonical ensemble ($\mu VT$-ensemble) and $NPT$-ensemble. Our algorithm is similar to Gibbs ensemble technique~\cite{Frenkel:UnderstandingMolecularSimulation2002}, which is used for simulating phase equilibria in $NVT$-ensemble. In one simulation step of 3-reservoirs method, the following processes are performed.
\begin{enumerate}[i)]
	\item with probability $p_G / 2$, trial particle insertion into the system
	\item with probability $p_G / 2$, trial particle deletion from the system
	\item with probability $p_V$, trial system size change
	\item with probability $1 - p_G - p_V$, trial displacement of one particle
\end{enumerate}
are chosen, where $p_G$ and $p_V$ are constants fixed in intervals $0 \le p_G, p_V \le 1$.

Via conventional algorithms of trial particle insertion and deletion in $\mu VT$-ensemble~\cite{Frenkel:UnderstandingMolecularSimulation2002}, we perform steps i) and ii), during which the system size is fixed. Steps i) and ii) satisfy the detailed balance condition~\cite{NorizoeForthcomingPaper:2010}, since this condition is, due to the consistency between the Boltzmann factors of the present system and $\mu VT$-ensemble, consistent with the condition in $\mu VT$-ensemble which has already been confirmed~\cite{Frenkel:UnderstandingMolecularSimulation2002}.

Via conventional algorithms in $NPT$-ensemble~\cite{Frenkel:UnderstandingMolecularSimulation2002}, step iii) is performed, during which $N$ is fixed. Step iii) also satisfies detailed balance due to the consistency between the Boltzmann factors of the present system and $NPT$-ensemble~\cite{NorizoeForthcomingPaper:2010}. However, unlike conventional Monte Carlo (MC) simulations in $NPT$-ensemble based on McDonald's method~\cite{Frenkel:UnderstandingMolecularSimulation2002}, $L_x, L_y$, and $L_z$ are independently changed in our 3-reservoirs method.

By Metropolis algorithm in $NVT$-ensemble, step iv) runs.

Here we discuss detailed balance and ergodicity, \textit{i.e.} statistical nature, of 3-reservoirs method. Steps i) and ii) change $N$ according to the detailed balance condition which corresponds to $\mu VT$-ensemble. Step iii) varies the system size according to the detailed balance condition corresponding to $NPT$-ensemble. The particle coordinates are updated in step iv), according to the standard Metropolis algorithm. As a result, 3-reservoirs method fulfills detailed balance and ergodicity, based on the conventional ensembles which satisfy ergodicity. This means $N$, $\left( L_x, L_y, L_z \right)$, and the particle coordinates are, in the phase space, simultaneously searched and tuned to the equilibrium state. This also indicates that statistics of $\mu PT$-ensemble contradicts none of the three underlying ensembles with 2 reservoirs, which is consistent with the thermodynamic consideration.

Maximization of statistical entropy per volume in $\mu PT$-ensemble corroborates the above results of statistics. The statistical entropy is defined as $-k_B \sum_j p_j \ln p_j$~\cite{Sack:1959,Reichl:ModernCourseInStatisticalPhysics}, where the suffix, $j$, denotes microstates of the system and $p_j$ is corresponding occurence probability. Due to indetermination of extensive variables in $\mu PT$-ensemble, we take volume density of extensive variables. The statistical entropy per volume is,
\begin{equation}
\label{eq:EntropyPerVolume}
	-k_B \sum\nolimits_j \frac{ p_j \ln p_j }{ V_j }.
\end{equation}
The probability distribution, $\{ p_1, p_2, \dots, p_j, \dotsc \}$, which maximizes eq.~\eqref{eq:EntropyPerVolume}, is determined under constraints:
\begin{gather}
	\label{eq:EntropyMaxConstraintsNormalization}
	\sum\nolimits_j p_j = 1,  \\
	\label{eq:EntropyMaxConstraintsEnergyAndParticleDensity}
	\sum\nolimits_j p_j \left( E_j \middle/ V_j \right) = \left\langle E \middle/ V \right\rangle,  \quad
	\sum\nolimits_j p_j \left( N_j \middle/ V_j \right) = \langle \rho \rangle,
\end{gather}
where $\langle \dotsm \rangle$ denotes the thermal average that should be specified by the reservoirs. Equation~\eqref{eq:EntropyMaxConstraintsNormalization} represents normalization condition. Equation~\eqref{eq:EntropyMaxConstraintsEnergyAndParticleDensity} comes from the thermodynamic degrees of freedom, which equal 2. Using Lagrange multipliers, we can solve this maximization problem~\cite{Reichl:ModernCourseInStatisticalPhysics}. The solution is~\cite{NorizoeForthcomingPaper:2010}:
\begin{gather}
\label{eq:StatisticalProbabilityDensityMuPT}
	p_j = e^{-1} \exp \left[ - \left( P \middle/ k_B T \right) V_j - \left( 1 \middle/ k_B T \right) E_j + \left( \mu \middle/ k_B T \right) N_j \right],  \\
\label{eq:EntropyPerVolumeAsThermalAverage}
	S / V = (1 / T) \left\{ P + \left\langle E / V \right\rangle - \mu \left\langle N / V \right\rangle + k_B T \left\langle 1 / V \right\rangle \right\}.
\end{gather}
$p_j$ is equivalent to the partition function eq.~\eqref{eq:SingleComponentPartitionFunctionmuPT}. The thermodynamic potential of $\mu PT$-ensemble,
\begin{equation}
\label{eq:ThermodynamicPotentialMuPT}
	\Phi = E- TS - \mu N + PV,
\end{equation}
is also obtained~\cite{NorizoeForthcomingPaper:2010}: $\Phi = -k_B T \ln e$, which is essentially zero compared with the other extensive variables in eq.~\eqref{eq:ThermodynamicPotentialMuPT} in the thermodynamic limit. This result is consistent with the Euler equation in thermodynamics. Moreover, the last term of eq.~\eqref{eq:EntropyPerVolumeAsThermalAverage} monotonically decreases with increasing $V$ for finite $V$. This means that the principle of maximizing entropy restricts the system to the finite size for computer simulations that treat finite system size. These results are also confirmed by maximization problems of $S/N$ and $S/E$.

Assuming that the ensemble averages of extensive variables, e.g. $N$ and $E$, were determined, Guggenheim formally introduced Boltzmann factor (statistical weight) of $\mu PT$-ensemble, based on analogy between other conventional ensembles~\cite{Guggenheim:1939,Koper:1996}. However, Guggenheim's assumption corresponds to, in the above calculation of the maximization problem, keeping the averages $\langle E \rangle$ and $\langle N \rangle$ fixed, instead of eq.~\eqref{eq:EntropyMaxConstraintsEnergyAndParticleDensity}. This contradicts the indetermination of extensive variables, as was pointed out by Prigogine and Sack~\cite{Prigogine:1950,Sack:1959}. Prigogine showed that the resulting partition function diverges and therefore concluded that no physical meaning is found in this partition function~\cite{Prigogine:1950,Sack:1959}. Free energy, $\Phi$, statistically determined via such partition function could be indefinite. Furthermore, in thermodynamics, $\Phi$ identically equals zero. Physical quantity that dominates $\mu PT$-ensemble, as corresponds to $F$ in $NVT$-ensemble, has been veiled since these early works.

On the other hand, in the present work, eq.~\eqref{eq:StatisticalProbabilityDensityMuPT} and $\Phi$ under the constraints eq.~\eqref{eq:EntropyMaxConstraintsEnergyAndParticleDensity} indicate the partition function equal to $e$. This $\Phi$ is smaller than the other extensive variables in eq.~\eqref{eq:ThermodynamicPotentialMuPT} and vanishes in thermodynamic limit. Moreover, we have found, in discussion on thermodynamics, that $NPT$, $\mu VT$, and $\mu PS$-ensembles underlie $\mu PT$-ensemble and that the 3 corresponding free energy densities of these underlying ensembles, rather than $\Phi$, are simultaneously minimized. This corresponds to the minimization of $F$ in $NVT$-ensemble.

As an example of 3-reservoirs method, we simulate equilibrium states of model polymer-grafted colloids. The model colloids, $\sigma_1$ in diameter, are interacting via spherically symmetrical repulsive square-step potential with a rigid core~\cite{Norizoe:2005}:
$\phi (r) = \infty \; (r < \sigma_1)$,
$\phi (r) = \epsilon_0 \; (\sigma_1 < r < \sigma_2)$,
$\phi (r) = 0 \; (\sigma_2 < r)$.
The distance between centers of the particles is denoted by $r$ and positive constants $\sigma_2$ and $\epsilon_0$ are the diameter and the height of the repulsive step resulting from the polymers grafted onto the colloidal particles. Simulating particles interacting via $\phi (r)$ in $NVT$-ensemble, we have studied phase behavior of these colloidal systems~\cite{Norizoe:2005}. These MC simulation results show that, at low $T$, high $P$, and $\sigma_2 / \sigma_1 \approx 2$, our particles self-assemble into string-like assembly, an amorphous solid. This assembly has also experimentally been observed~\cite{Osterman:2007}.

In these recent studies at finite $T$ in both 2 and 3-dimensions, the string-like assembly has been found to be metastable, locally directed in the same direction and globally-isotropic, in $NVT$-ensemble. Equilibrium states at finite $T$ are still open, though various ground states of our model at zero $T$ have been found via genetic algorithms~\cite{Pauschenwein:2008}.

Using 3-reservoirs method, at finite $T$, we simulate equilibrated states of our model system in 2-dimensions. $\sigma_1$ and $\epsilon _0$ are taken as unit length and unit energy respectively. Dimensionless chemical potential is defined as~\cite{DoctoralThesis},
\[
	\mu' := \mu / k_B T - \ln \left( \left. \varLambda^2 \right/ \sigma_1^2 \right) = \ln \left( \rho_{\text{res}}^{(1)} \, \sigma_1^2 \right), \quad \varLambda = \frac{ h }{ \sqrt{2 \pi m k_BT} },
\]
where $h$ denotes Planck's constant, and $m$ mass of the particle.

In the initial state, $N_0 = 1254$ particles are arranged on homogeneous triangular lattices in a square system box with the periodic boundary condition. Initially, $\rho \sigma_1^2 = 0.451$. In step iv), a particle is given uniform random trial displacement within a square $0.4 \sigma_1$ long in each direction. In step iii)~\cite{NorizoeForthcomingPaper:2010}, $(L_x, L_y)$ is changed to $( \, L'_x = L_x + \varDelta L (1 - 2 \xi_x), \quad L'_y = L_y + \varDelta L (1 - 2 \xi_y) \, )$,
where $\varDelta L$ is constant length, fixed at $0.01 \sigma_1$, and $\xi_x$ and $\xi_y$ are random numbers uniformly distributed over intervals $0 \le \xi_x, \xi_y \le 1$. $p_G = 10 \%$ and $p_V = 1 / N_0$. With this $p_V$, computational time is about twice as long as simulation in $NVT$-ensemble. In the following, we define 1 Monte Carlo step (MCS) as $N_0$ simulation steps. $\sigma_2 / \sigma_1 = 2.0$ is fixed.

\begin{figure*}[!tb]
	\centering
	\includegraphics[clip]{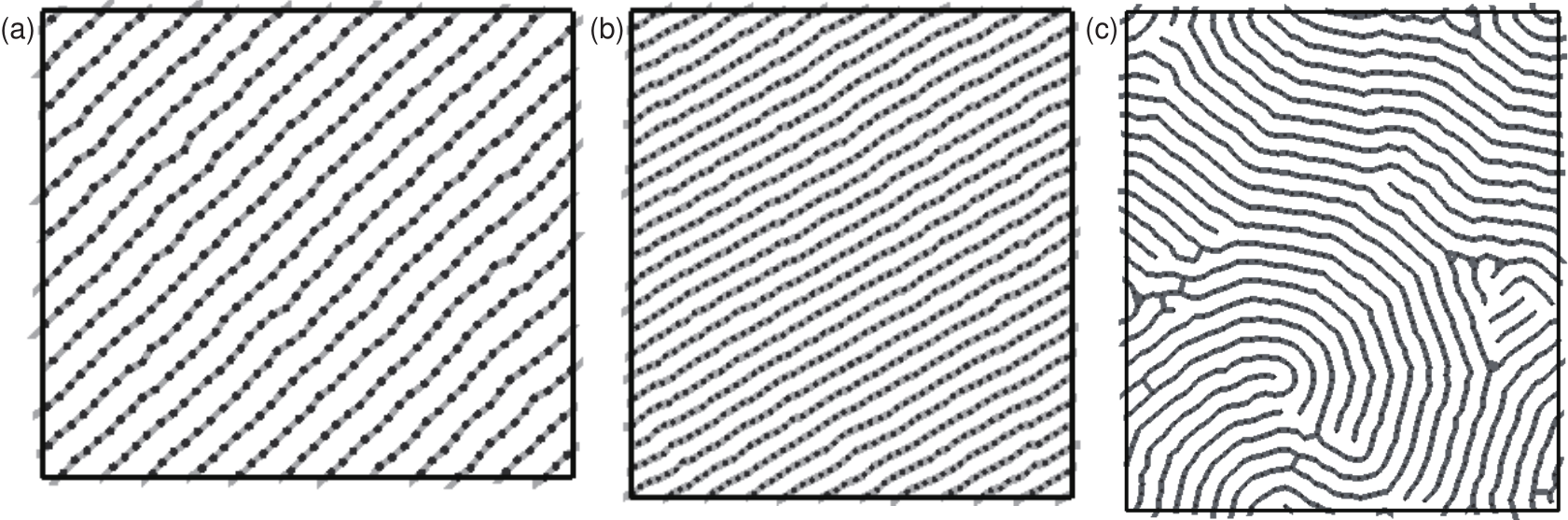}
	\caption{Snapshots of the system.
	(a), $k_B T / \epsilon_0 = 0.12 $, $P \sigma_1^{2} / \epsilon_0 = 1 $ and $\mu' = 27.63 $. 
	(b), $k_B T / \epsilon_0 = 0.12 $, $P \sigma_1^{2} / \epsilon_0 = 1.1 $ and $\mu' = 29.93 $. 
	Black dots represent the centers of the particles and grey lines denote networks of overlaps between the particles. A snapshot of the system simulated in $NVT$-ensemble at $k_B T / \epsilon_0 = 0.12 $, $\rho \sigma_1^2 = 0.451$ and $N = 1200$ is also shown in (c)~\cite{Norizoe:2005}.}
	\label{fig:Frog2DmuPTS20Snapshots}
\end{figure*}

\begin{figure}[!tb]
	\centering
	\includegraphics[clip]{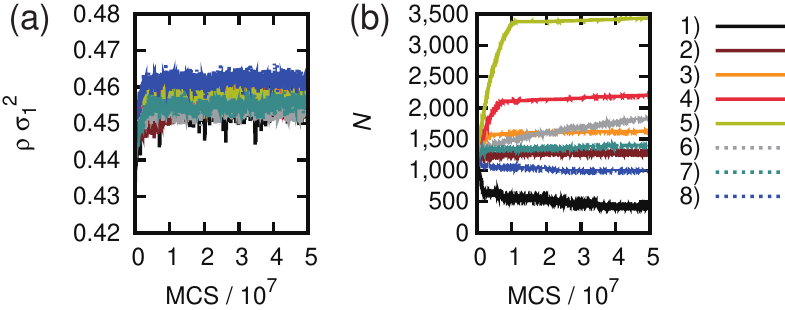}
	\caption{Time evolution of $\rho$, (a), and time evolution of $N$, (b), sampled at $k_B T / \epsilon_0 = 0.12$ and various $P \sigma_1^{2} / \epsilon_0$, $\mu'$. Lines 1) to 5) are results at $P \sigma_1^{2} / \epsilon_0 = 1.0$.
		1): 
			$\mu' = 27.63$.
		2): 
			$\mu' = 29.93$.
		3): 
			$\mu' = 31.54$.
		4): 
			$\mu' = 32.23$.
		5): 
			$\mu' = 34.53$.
		Lines 6) to 8) are results at $\mu' = 29.93$.
		6): $P \sigma_1^{2} / \epsilon_0 = 0.95$. 
		7): $P \sigma_1^{2} / \epsilon_0 = 0.98$. 
		8): $P \sigma_1^{2} / \epsilon_0 = 1.1$. 
	}
	\label{fig:Frog2DmuPTS20T012ForArticle-RhoAndN}
\end{figure}

First, simulation results at low temperature, $k_B T / \epsilon_0 = 0.12$, are discussed. In $NVT$-ensemble at this $k_B T / \epsilon_0$, the string-like assembly is observed in regions $0.412 \lessapprox \rho \sigma_1^2 \lessapprox 0.481$ and the string length diverges at $\rho \sigma_1^2 \approx 0.451$~\cite{Norizoe:2005}. Snapshots of the system in the present simulation are shown in Figs.~\ref{fig:Frog2DmuPTS20Snapshots}(a) and (b). Despite the different parameters, both the snapshots present globally-anisotropic defect-free string-like assembly, although a snapshot of the system simulated in $NVT$-ensemble at $\rho \sigma_1^2 = 0.451$ and $N = 1200$, shows, in Fig.~\ref{fig:Frog2DmuPTS20Snapshots}(c)~\cite{Norizoe:2005}, globally-isotropic string-like assembly. Time evolution of $\rho$ and $N$ are plotted in Fig.~\ref{fig:Frog2DmuPTS20T012ForArticle-RhoAndN}. The same assembly is also observed at all the various $(\mu', P \sigma_1^{2} / \epsilon_0)$ chosen in Fig.~\ref{fig:Frog2DmuPTS20T012ForArticle-RhoAndN}. Time evolution of $\rho$, Fig.~\ref{fig:Frog2DmuPTS20T012ForArticle-RhoAndN}(a), all fluctuates in the vicinity of $\rho \sigma_1^2 \approx 0.451$, regardless of $(\mu', P \sigma_1^{2} / \epsilon_0)$. These results indicate these parameter sets $(\mu', P \sigma_1^{2} / \epsilon_0)$ are located in the vicinity of the same TSP, and that the system reaches the identical equilibrium state. Simulations started from different initial conditions, \textit{e.g.} different initial $\rho$ and $L_x / L_y$, also reach these results. Simulations resumed from the instantaneous state of Fig.~\ref{fig:Frog2DmuPTS20Snapshots}(a) by disconnecting reservoirs 1 or 2, \textit{i.e.} simulations resumed in $NPT$ or $\mu VT$-ensembles, keep the same assembly and $\rho$ (data not shown), which corroborates our heuristic method for determining TSP. On the other hand, time evolution of $N$, Fig.~\ref{fig:Frog2DmuPTS20T012ForArticle-RhoAndN}(b), shows that the total system size depends on $(\mu', P \sigma_1^{2} / \epsilon_0)$. $N$ and $V$ rise with large $\mu'$ and small $P \sigma_1^{2} / \epsilon_0$, whereas, regardless of the chosen $(\mu', P \sigma_1^{2} / \epsilon_0)$, computationally handy $N$ and $V$ are kept for a long time, within which good statistics of simulation results are obtained. With such results that are accurate enough to keep the equilibrated structure of the system in $\mu PT$-ensemble, we are allowed to switch the ensemble to conventional ones, \textit{e.g.} $NPT$ or $NVT$-ensembles, and to perform long simulation runs, free from the divergence and the vanishment of $N$, of the equilibrated structure obtained via the 3-reservoirs method. The system outside the chosen $(\mu', P \sigma_1^{2} / \epsilon_0)$, \textit{i.e.} outside the vicinity of TSP, diverges or vanishes quickly (data not shown), which facilitates heuristically determining TSP quickly. These results coincide with thermodynamics discussed in the introduction. Globally-anisotropic defect-free triangular crystals of the outer cores, $\sigma_2$, are also observed at this temperature (data not shown)~\cite{NorizoeForthcomingPaper:2010}, which shows our method is applicable to crystalline states. Since any crystals fit into rectangular system boxes with periodic boundary conditions, other shapes of system boxes, \textit{i.e.} a combination of Parrinello-Rahman technique~\cite{Frenkel:UnderstandingMolecularSimulation2002} and our method, are left for the future.

At $k_B T / \epsilon_0 = 1.5 $, $P \sigma_1^{2} / \epsilon_0 = 10 $ and $\mu' = 12.50 $, a disordered state is observed and at $k_B T / \epsilon_0 = 0.01 $, $P \sigma_1^{2} / \epsilon_0 = 1 $ and $\mu' = 34.53 $, a triangular crystal (data not shown).


In conclusion, we have shown that $\mu PT$-ensemble is built by combining $NPT$, $\mu VT$, and $\mu PS$-ensembles. The 3 corresponding free energy densities are simultaneously minimized in $\mu PT$-ensemble, rather than $\Phi$ is. Unlike the early works shown above, thermodynamics outside TSP and maximization of $S/V$ have also been discussed. We have devised 3-reservoirs method and simulated the colloidal suspension which are interacting via hard core with step repulsive potential. Unlike other advanced techniques, our method allows tracing the physical trajectories in the phase space and quickly starting production runs without advanced programming and large amounts of preliminary simulation. This facilitates and reduces the total work flow of simulation studies, including data analysis. Simulation results coincide with thermodynamics. Simultaneously tuning $N$ and the system size, which is feasible only via 3-reservoirs method, and crossing non-equilibrium, the systems reach equilibrium at low $T$. These results illustrate our method is applicable to ordered equilibrium states of various physical systems at finite $T$. This unique advantage, \textit{i.e.} the simultaneous tuning of $N$ and the system size, of 3-reservoirs method could overturn previous simulation results obtained via the other simulation techniques. The equation of state linking $\mu$, $P$, and $T$ is also calculated.



\begin{acknowledgments}
The authors wish to thank Professor Komajiro Niizeki and Mr Masatoshi Toda for helpful suggestions and discussions.
This work is partially supported by a grant-in-aid for science ``Soft Matter Physics'' from the Ministry of Education, Culture, Sports, Science, and Technology, Japan.
\end{acknowledgments}


%

\end{document}